\documentclass[prb,aps,twocolumn,superscriptaddress,showpacs,footinbib,longbibliography]{revtex4-1}
\usepackage[colorlinks=true,linkcolor=black, citecolor=blue, urlcolor=blue, 
unicode=true,breaklinks]{hyperref}

\usepackage{graphicx,amsmath,braket,tikz,amssymb,footmisc,lipsum,array,tabularx,float,comment,xcolor}
\usepackage{epstopdf}
\usepackage{amsmath}
\usepackage{dsfont} 
\usepackage{pgfplots}
\usepackage{physics}
\usepackage{cleveref}

\begin{document}
	
\title{Magnetic field tuned superconducting and normal phase magnetism in CeCo$_{0.5}$Rh$_{0.5}$In$_{5}$}
	
\author{A. Howell}
\affiliation{School of Physics and Astronomy, University of Edinburgh, Edinburgh EH9 3JZ, UK}
\author{M. Songvilay}
\affiliation{School of Physics and Astronomy, University of Edinburgh, Edinburgh EH9 3JZ, UK}
\author{J. A. Rodriguez-Rivera}
\affiliation{NIST Center for Neutron Research, National Institute of Standards and Technology, 100 Bureau Dr., Gaithersburg, MD 20899}
\affiliation{Department of Materials Science, University of Maryland, College Park, MD  20742}
\author{Ch. Niedermayer}
\affiliation{Laboratory for Neutron Scattering, Paul Scherrer Institut, CH-5232 Villigen, Switzerland}
\author{Z. Husges}
\affiliation{Helmholtz-Zentrum Berlin für Materialien und Energie, 14109 Berlin, Germany}
\author{P. Manuel}
\affiliation{ISIS Facility, Rutherford Appleton Laboratory, Chilton, Didcot OX11 0QX, United Kingdom}
\author{S. Saha}
\affiliation{Center for Nanophysics and Advanced Materials, Department of Physics, University of Maryland, College Park, Maryland 20742, USA}
\author{C. Eckberg}
\affiliation{Center for Nanophysics and Advanced Materials, Department of Physics, University of Maryland, College Park, Maryland 20742, USA}
\author{J. Paglione}
\affiliation{Center for Nanophysics and Advanced Materials, Department of Physics, University of Maryland, College Park, Maryland 20742, USA}
\author{C. Stock}
\affiliation{School of Physics and Astronomy, University of Edinburgh, Edinburgh EH9 3JZ, UK}
	
\date{\today}
	
\begin{abstract}
		
By tuning the superconducting order parameter with an applied magnetic field, we use neutron diffraction to compare the magnetic ordered phases in superconducting and normal states of CeCo$_{0.5}$Rh$_{0.5}$In$_{5}$.  At zero applied field, CeCo$_{0.5}$Rh$_{0.5}$In$_{5}$ displays both superconductivity ($T_{c}$=1.3 K) and spatially long-ranged commensurate $\uparrow\downarrow\uparrow\downarrow$ antiferromagnetism ($T_{N}$=2.5 K, propagation vector $\vec{Q}=({1\over 2}, {1\over 2}, {1\over 2})$). Neutron spectroscopy fails to measure propagating magnetic excitations from this ground state with only temporally overdamped fluctuations observable.  On applying a magnetic field which suppresses superconductivity, we find anisotropic behavior in the static magnetism.  When the applied magnetic field is along the crystallographic $c$-axis, no change in the static magnetic response is observable.  However when the field is oriented within the $a-b$ plane, an increase in $T_{N}$ and change in the critical response are measured.  At low temperatures in the superconducting phase, the elastic magnetic intensity increases linearly ($\propto |H|$) with small $a-b$ oriented fields.  However, this trend is interrupted at intermediate fields where commensurate block $\uparrow\uparrow\downarrow\downarrow$ magnetism with propagation vector $\vec{Q}=({1\over 2}, {1\over 2}, {1\over 4})$ forms. For large applied fields in the [1 $\overline{1}$ 0] direction which completely suppresses superconductivity, weakly incommensurate magnetic order along $L$ is observed to replace the commensurate response present in the superconducting and vortex phases. We discuss experimental considerations related to this shift in momentum and suggest field-induced incommensurate static magnetism, present in the normal state of superconducting and antiferromagnetic CeCo$_{0.5}$Rh$_{0.5}$In$_{5}$ for $a-b$ plane oriented magnetic fields.  We speculate that these field dependent properties are tied to the field induced anisotropy associated with the local Ce$^{3+}$ crystal field environment of the tetragonal `115' structure. 
		
\end{abstract}
	
	
\maketitle

\section{Introduction}

While localized magnetism maybe viewed as incompatible with itinerant superconductivity, work in the past several decades has found a number of systems where superconductivity is proximate to static magnetic phases.~\cite{Miyake86:34,Norman11:332}  Notable examples with large superconducting transition temperatures include the cuprate~\cite{Birgeneau06:75,Norman03:66,Kastner98:70}, iron (pnictide and chalcogenide) based compounds~\cite{Dai15:87}, \textcolor{black} {and nickel based compounds including recent work under pressure~\cite{Sun23:621} and previous work on layered borocarbides~\cite{Pickett94:72,Goldman94:50,Canfield98:51,Lynn97:55,Muller01:64}.}  These materials display a delicate interplay between magnetic and superconducting phases~\cite{Lee06:78,Kivelson03:75} and investigating these compounds near the boundary between magnetism and superconductivity often results in the discovery of new phases.   For example, the use of high magnetic fields to suppress cuprate superconductivity has led to the discovery of intertwined orders including a charge density wave that competes with high temperature superconductivity.~\cite{Chang12:8,Fradkin15:87}  

The `115' series of compounds is based on chemically substituted and doped variants of CeCoIn$_{5}$ with phase diagrams displaying analogous interplay between localized magnetism and superconductivity to that found in cuprate and iron-based compounds.  CeCoIn$_{5}$ is an unconventional superconductor with the highest transition temperature ($T_{c}$=2.3 K) recorded for heavy fermion materials~\cite{Petrovic01:13,Stewart84:56,Pfleiderer09:81,Thompson12:81} and has a $d$-wave superconducting order parameter.~\cite{Izawa01:87,Aoki01:87}  The crystallographic structure is based on a tetragonal unit cell with layers of magnetic Ce$^{3+}$ stacked along $c$. 

In terms of the magnetic response from the Ce$^{3+}$ ions in CeCoIn$_{5}$, the high temperature non superconducting (normal) state consists of overdamped magnetic excitations peaked near $\vec{Q}=({1\over 2}, {1\over 2}, {1\over 2})$, indicative of antiferromagnetic interactions between the Ce$^{3+}$ ions, both within the $a-b$ plane and along $c$. We note that no static magnetism at zero applied field has been reported.  However, in the superconducting phase, the momentum broadened fluctuations are replaced by a temporally sharp resonance peak~\cite{Stock08:100,Stock12:109_2,Raymond15:115,Song16:7} characterized by fluctuations polarized along the $c$-axis.  This illustrates a link between superconductivity and localized magnetism in the `115' series of compounds. 

The fluctuating magnetism in superconducting CeCoIn$_{5}$ contrasts with static antiferromagnetism in CeRhIn$_{5}$~\cite{Fobes17:29}.  Static incommensurate magnetism ($T_{N}$=3.8 K) (propagation wave vector $\vec{Q}=({1\over 2}, {1\over 2}, 0.297)$) representative of helical magnetic order along the crystallographic $c$-axis is found based on unpolarized~\cite{Bao00:62} and, required given the symmetry of the structural unit cell, polarized neutron diffraction. The excitations in CeRhIn$_{5}$ are dominated by strong transverse spin waves which are well defined and underdamped in energy.~\cite{Das14:113, Fobes17:29}  However, the transverse in-plane magnetic fluctuations are unstable, displaying a breakdown into multiparticle states with increased energy scale.~\cite{Stock15:114,Brener24:110} In-plane excitations are also not observable when CeCoIn$_{5}$ is tuned towards antiferromagnetism with Hg substitution~\cite{Stock18:121}, despite the coexistence of static commensurate (propagation wave vector $\vec{Q}=({1\over 2}, {1\over 2}, {1\over 2})$) magnetism.  

\begin{figure}
	\includegraphics[width=100mm,trim=11cm 2.75cm 9.5cm 2cm,clip=true]{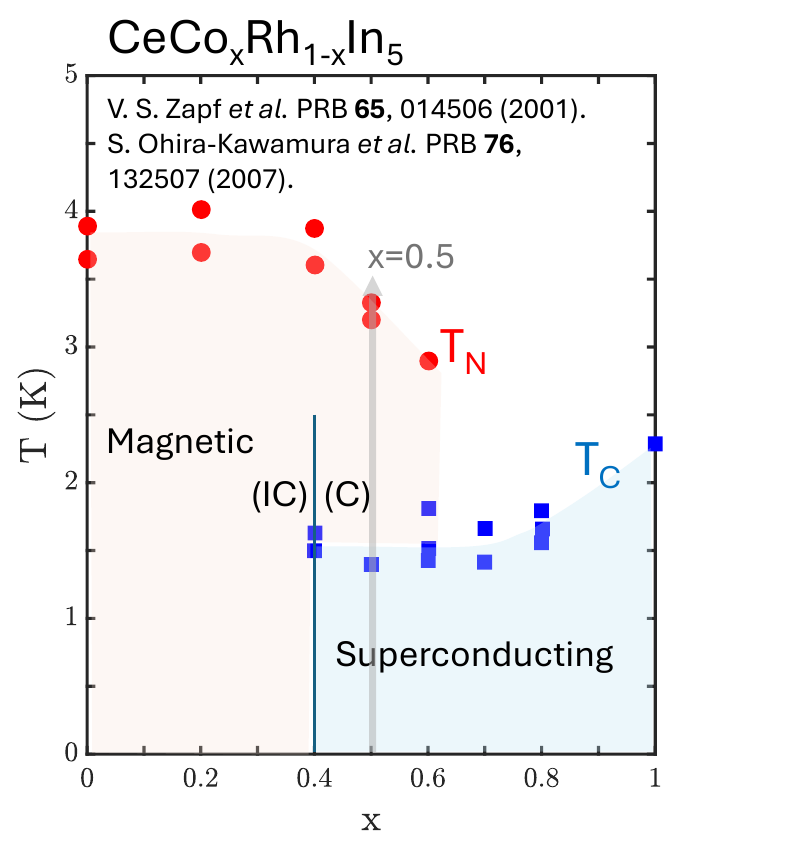}
	\caption{The magnetic and superconducting phase diagram of CeCo$_{x}$Rh$_{1-x}$In$_{5}$ adapted from Refs. \onlinecite{Zapf01:65} and \onlinecite{Kawamura07:76}.  The data is a compilation of transition temperatures measured with heat capacity, resistivity, and AC/DC susceptibility.  The position of our CeCo$_{0.5}$Rh$_{0.5}$In$_{5}$ sample with respect to the entire phase diagram is illustrated by the vertical grey arrow.} \label{fig:phase_diagram}
\end{figure}

While a number of chemical substitution and doping techniques illustrate an interplay between static magnetism and superconductivity, in this study we focus on the specific case of CeCo$_{x}$Rh$_{1-x}$In$_{5}$.  The electronic and magnetic phase diagram of CeCo$_{x}$Rh$_{1-x}$In$_{5}$ is displayed in Fig. \ref{fig:phase_diagram} with data points taken from Refs. \onlinecite{Zapf01:65} and \onlinecite{Kawamura07:76}.  In terms of the static magnetic order, for small concentrations of Co substitution defined by the parameter $x$ (on the horizontal axis in Fig. \ref{fig:phase_diagram}), incommensurate magnetic order~\cite{Bao00:62,Llobet04:69,Hegger00:84,Knebel06:74,Kawasaki03:91,Muramatsu01:70,Park08:105,Chen06:97,Paglione08:77} is present (as discussed in the case of CeRhIn$_{5}$ above) and on approaching superconducting concentrations near $x$ $\sim$ 0.4, becomes commensurate~\cite{Yokoyama06:75}.  Noteably, at the concentration $x$=0.4 studied in Ref. \onlinecite{Yokoyama06:75}, both incommensurate and commensurate magnetic order are found to coexist.  We note that the exact reported concentration of $x$ where superconductivity is onset varies from $x$ $\sim$ 0.35-0.4~\cite{Zapf01:65,Kawamura07:76} and in Fig. \ref{fig:phase_diagram} we have taken the data points from Ref. \onlinecite{Zapf01:65}.  Through Co$\leftrightarrow$Rh subsitution, static magnetic and superconducting phases can be tuned.

As illustrated in Fig. \ref{fig:phase_diagram} the Co concentration of $x\sim 0.4$, separates two distinct phases.  The electronic Fermi surface evolution across this boundary has been characterized by quantum oscillations~\cite{Goh08:101} and found to be marked by an abrupt change in the Fermi surface and also the associated effective mass.  For comparison purposes, the position of the concentration studied in our current paper with neutrons is illustrated by the vertical arrow at $x$=0.5 in Fig. \ref{fig:phase_diagram}.  Our sample is near the critical Co concentration of $x\sim$0.4 yet displays both commensurate antiferromagnetism ($T_{N}$=2.5 K) and a low temperature superconductivity ($T_{c}$=1.3 K).~\cite{Zapf01:65,Pagliuso01:64,Pagliuso02:312}

Given that superconducting states consist of fluctuating variants of the parent phases~\cite{Lee06:78,Kivelson03:75},  studying competing phases in compounds near the onset to superconductivity will provide information on these fluctuations and energy scales important to superconductivity.  One means of doing this is to apply strong magnetic fields at low temperatures to suppress the superconducting order parameter to reveal the underlying competing states in the normal state.  This is difficult in the cuprate or iron based unconventional superconductors owing to the large magnetic fields required to overcome the critical $H_{c2}$ field where superconductivity is suppressed.  The CeCo$_{x}$Rh$_{1-x}$In$_{5}$ (or more broadly the `115' series) compounds displays unconventional superconducting order parameters, but with reduced transition temperatures and energy scales that are more amenable to experiments.  In this work, we focus on the CeCo$_{0.5}$Rh$_{0.5}$In$_{5}$ concentration which is located near the boundary between incommensurate antiferromagnetism and superconductivity and displays both static magnetism and low temperature superconductivity as discussed above.  We characterize the excitations and apply a magnetic field to suppress the superconducting order parameter to reveal the underlying competing magnetic phases which we study with neutron diffraction.

\section{Experiments}

The \textcolor{black} {$\sim$ 1.5 g} sample of CeCo$_{0.5}$Rh$_{0.5}$In$_{5}$ was grown using the flux technique.  The sample has a superconducting transition temperature of $T_{c}$=1.3 K and N\'eel order at $T_{N}$=2.5 K.  \textcolor{black}{A piece of this sample was used in Ref. \onlinecite{Goh08:101} with heat capacity displaying two peaks shown in Fig. 1. We discuss homogeneity of Co $\longleftrightarrow$ Rh substitutes in the context of the magnetic transitions measured with neutrons below.} $H_{c2}$ of our sample was measured to be 7.5 T when the field is oriented within the $a-b$ plane.

Neutron diffraction experiments were performed using the RITA2 (PSI, Villigen, Switzerland) and FLEXX (HZB, Berlin, Germany) cold triple-axis spectrometers and at the WISH diffractometer (ISIS, Didcot, UK) with Bragg reflections of the form (H, H, L) aligned within the horizontal scattering plane.  For RITA2, the incident and final energies were fixed to E$_{f}$=3.5 meV, with a Beryllium filter placed after the sample.  On FLEXX, an incident energy of E$_{i}$=3.0 meV was used. 

Several different magnetic field geometries were used for this work requiring multiple instruments and configurations.  Because of kinematic constraints of neutron scattering and the requirement to study momentum transfers that reached magnetic Bragg positions of the form (${1\over 2}$,${1\over 2}$,L) (where $L\neq 0$), the sample needed to be aligned such that Bragg peaks of the form (HHL) lay within the $\textit{horizontal}$ scattering plane.  This meant that vertical magnetic fields (which provide maximum incident and scattered neutron beam access) were fixed along the $[1\overline{1}0]$ axis.  Vertical fields of $\mu_{0}H$ $<$ 9 T were available on RITA2 allowing us to obtain data both in the normal and superconducting phases of our sample.  To study the magnetic field dependence well above $H_{c2}$ in the normal phase, we used the vertical 14.5 T field available at FLEXX (HZB, Berlin).  Horizontal magnets allow the beam, in principle, to be applied along any crystallographic direction in the horizontal scattering plane defined by wavevectors of the form [HHL].  Therefore the horizontal magnetic field can be aligned both within the crystallographic $a-b$ plane (along [110], for example) or along the $c$ axis (along [001]) while allowing kinematic constraints of neutron scattering to be satisfied.  However, horizontal magnetic fields require multiple solenoids to allow access for incident and scattered neutron beams and are currently constrained to fields below $\mu_{0}H$ $<$ 6.5 T (available at, for example, RITA2, PSI) owing to engineering constraints of the strong stray fields.    Throughout the paper below we note which type of field and orientation is being used at the beginning of each result subsection.

We note that the field was always changed in the normal state at high temperatures.  To improve thermal conductivity of the sample mount, for zero magnetic field results reported the sample was always cooled in a 0.03 T magnetic field to suppress superconductivity of Aluminum.  This has been measured by us in previous low temperature studies of other heavy fermion materials~\cite{Stock12:109} to improve thermal conductivity and allow lower temperatures to be achieved at the sample position.

Given that one of the goals of this study is to compare the magnetism in antiferromagnetic and superconducting CeCo$_{0.5}$Rh$_{0.5}$In$_{5}$ and correlate the results with parent antiferromagnetic CeRhIn$_{5}$ and superconducting CeCoIn$_{5}$, we investigated the magnetic fluctuations from the magnetic commensurate static $\vec{Q}=({1\over 2},{1\over 2},{1\over 2})$ phase using neutron inelastic scattering. Neutron spectroscopy measurements were performed on the MACS triple-axis spectrometer at NIST (Gaithersburg, USA).~\cite{Rodriguez08:19}  Given the need to sample a large region of momentum space, the multi detector arrangement and large flux on the sample was advantageous. Cooled filters of Beryllium were used after the sample with the final energy fixed by a double bounce PG(002) analyzer with E$_{f}$ =5.1 meV and E$_{i}$ defined by a double focused PG(002) monochromator.  

\section{Results}

The main focus of this paper is magnetic neutron diffraction in a magnetic field and this is discussed below utilizing two cold neutron spectrometers.  We conclude the results section with a presentation of the magnetic excitations and compare these to previously published results in antiferromagnetic and non superconducting CeRhIn$_{5}$ and superconducting CeCoIn$_{5}$.

\begin{figure}[t]
	\includegraphics[width=90mm,trim=1.5cm 4.75cm 1.75cm 4.25cm,clip=true]{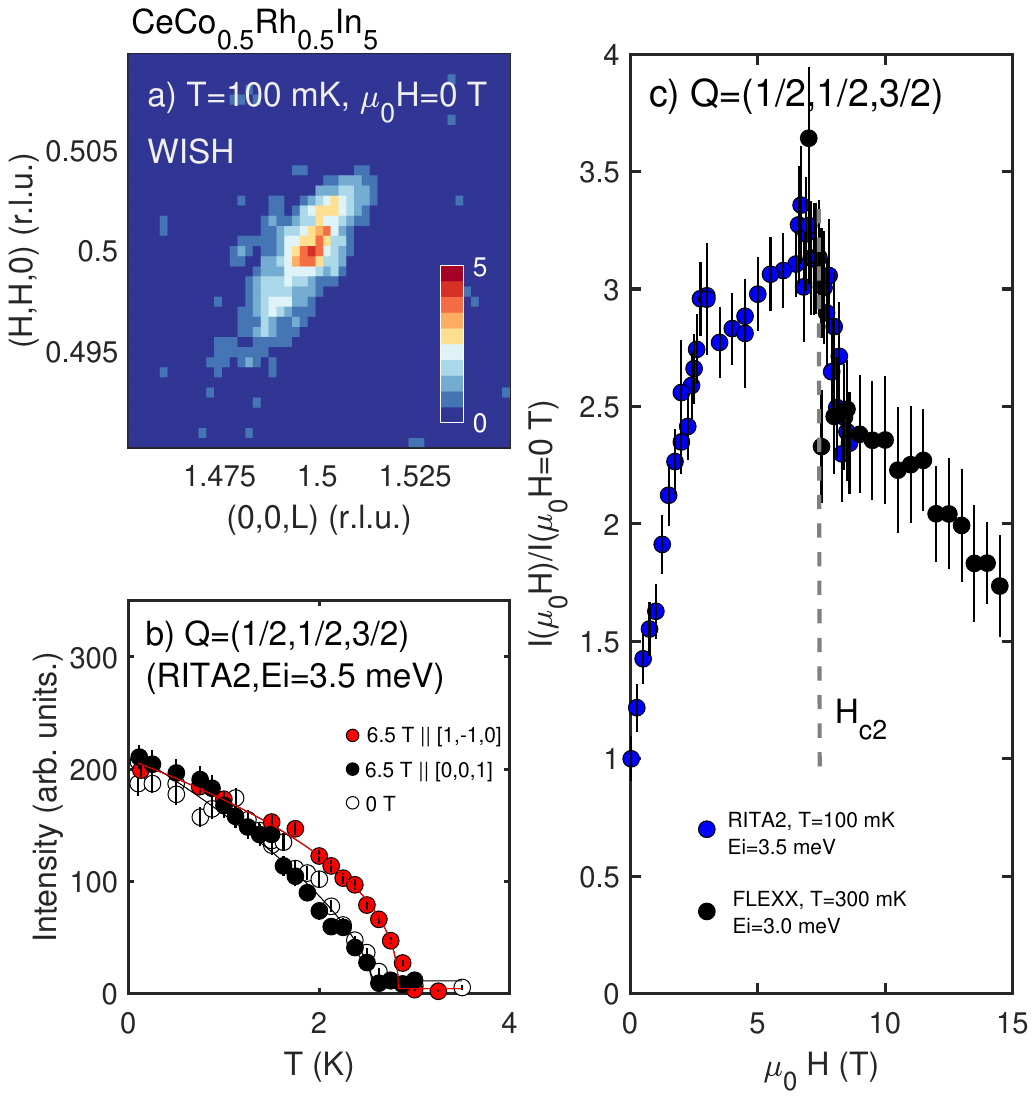}
	\caption{Temperature and applied field dependent magnetic diffraction with horizontal and vertical fields at $\vec{Q}_{0}=({1\over2},{1\over2},{3\over2})$.  $(a)$ illustrates a colormap of the magnetic Bragg peak at T=100 mK illustrating the $\vec{Q}$=(1/2, 1/2, 1/2) commensurate ordering.  $(b)$ The magnetic order parameter at zero and applied fields with field oriented along the [$\overline{1}$, 1, 0] and [0, 0, 1] axes measured using the RITA2 triple-axis spectrometer. \textcolor{black} {Note that the data for differing field orientations have been normalized to agree at the lowest temperatures so that order parameter curvatures can be compared directly.}  $(c)$ displays the \textcolor{black}{field dependence of the relative change (normzlized to zero field) in the magnetic Bragg peak intensity at T=100 mK (RITA2) and T=300 mK (FLEXX) (note that all values are taken with respect to $\mu_{0}$H=0 T data)}.  Note that the maximum field available on RITA2 was $\mu_{0}$H=9 T and 15 T on FLEXX so an overlap of 5-9 T is illustrated for data normalization.  H$_{c2}$ from susceptibility is shown.} \label{fig:fig1}
\end{figure}

We first discuss the static magnetic properties of CeCo$_{0.5}$Rh$_{0.5}$In$_{5}$ as a function of applied magnetic field and temperature which are summarized in Fig. \ref{fig:fig1} for the magnetic field applied along both the crystallographic $c$-axis and within the $a-b$ plane.  The temperature dependence of the commensurate $\vec{Q}=({1\over 2},{1\over 2},{1\over 2})$ (Fig. \ref{fig:fig1} $a$),  magnetic Bragg peak is illustrated in Fig. \ref{fig:fig1} $(b)$.  The magnetic nature of this Bragg peak is confirmed as it is absent at high temperatures.  The magnetic Bragg peak is centered around both the commensurate (within experimental resolution) H and L=${1\over2}$ and is indicative of a doubling of the magnetic unit cell consistent with antiferromagnetic correlations.  This forms a real-space $\uparrow\downarrow$ arrangement of spins along the crystallographic $c$ and within the $a-b$ plane.  The width of the magnetic peak is resolution limited both along the (H, H, 0) and (0, 0, L) directions representative of a real space correlation length of at least $\sim$ 500 \AA\, set by the experimental resolution of the triple-axis spectrometer.  The observed magnetic structure contrasts with the incommensurate order reported in non superconducting variants such as CeRhIn$_{5}$~\cite{Bao00:62,Fobes17:29,Stock15:114} discussed in the introduction.

\begin{figure}[t]
	\includegraphics[width=93mm]{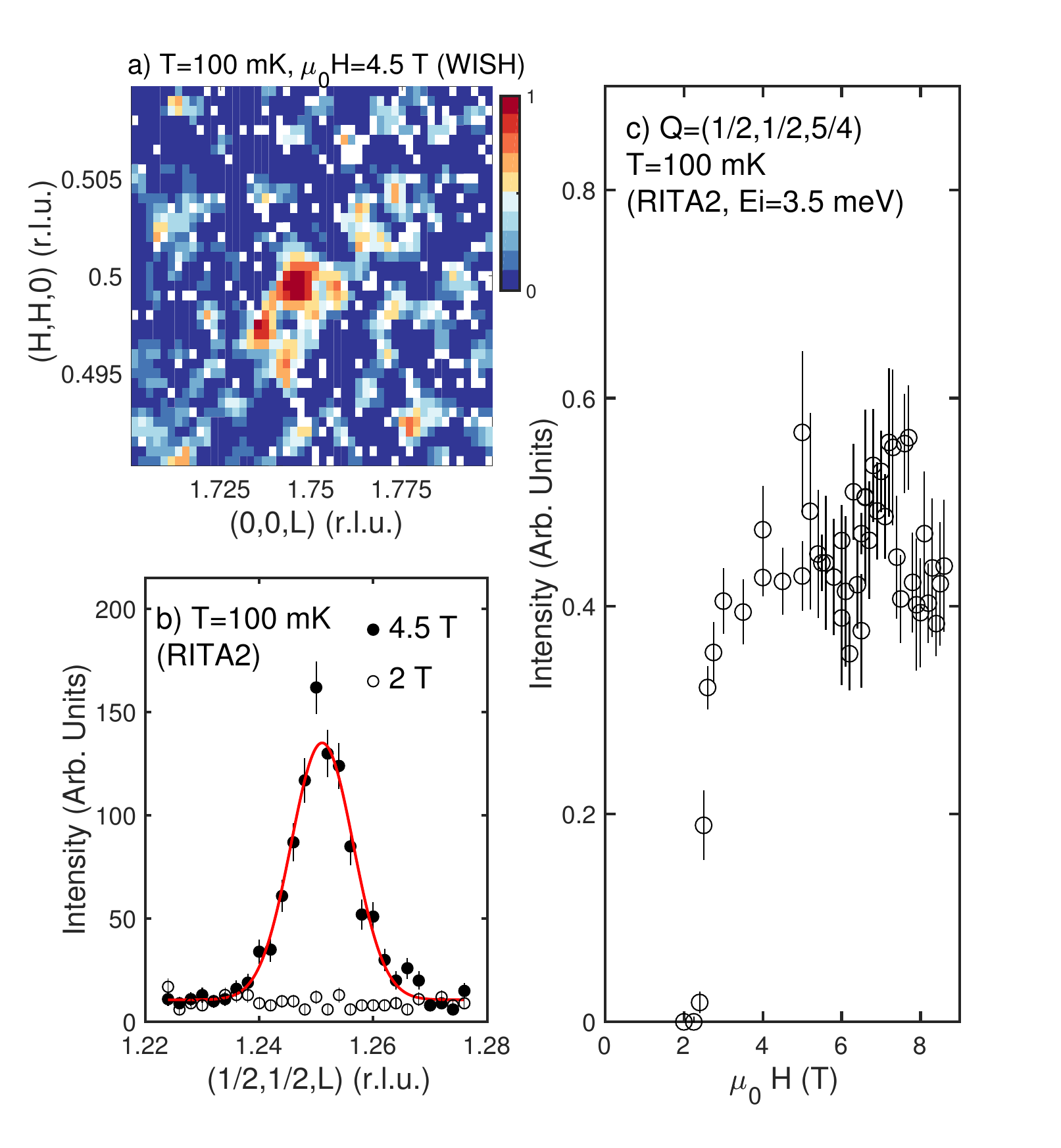}
	\caption{Magnetic diffraction in vertical fields at $\vec{Q}_{0}=({1\over 2},{1\over 2},{1\over 4})$.  $(a)$ illustrates a map in reciprocal space taken on WISH with $(b)$ displaying representative scans taken on RITA2 showing the presence of the L=1/4 type order with an applied field within the $a-b$ crystallographic plane.  $(c)$ The intensity of the magnetic Bragg peak at $\vec{Q}$=(${1\over 2}$, ${1\over 2}$, 1.25) as a function of applied magnetic field at T=100 mK taken on RITA2.  No L=${1\over 4}$ magnetic Bragg peaks were observed at fields up to $\mu_{0}H$=6.5 T with the field aligned along the $c$-axis.} \label{fig:fig2}
\end{figure}

\subsection{Anisotropic magnetic critical properties with applied horizontal and vertical magnetic fields}

The critical properties are now discussed which are indicative of the underlying anisotropy of the static magnetism as sampled from the temperature dependence of the magnetic order parameter as illustrated in Fig. \ref{fig:fig1}. $(a,b)$.  While the superconducting transition temperature does not vary significantly with Co $\longleftrightarrow$ Rh substitution (Fig. \ref{fig:phase_diagram}), the magnetic T$_{N}$ transition temperature does and \textcolor{black} {the sharp (in temperature) onset of magnetic order in our sample indicates homogeneous chemical substitution.}   The intensity of the magnetic Bragg peak is a measure of the magnetic order parameter and is fit to a power law in Fig. \ref{fig:fig1} $(b)$ $I(T) \propto |M(T)|^{2} \propto (T_{N}-T)^{2\beta}$.  Given that power law correlations are only expected at temperatures in the limit of $T \rightarrow T_{N}$, the value of the extracted critical exponent is naturally sensitive to the temperature region where the fit is performed.  We have fit power laws here to temperatures above 1 K and below $T_{N}$ given that power law correlations are only expected near $T_{N}$.  \textcolor{black} {We note that this fit provides agreement over a much broader temperature range and taking smaller temperature ranges gave unreliable fits to the exponents with large errorbars and fits over a broader temperature range did not change the exponent results within error.}  

With the temperature dependent magnetic intensity plotted in Fig. \ref{fig:fig1} $(b)$, we note that the magnetic intensity seems to continue increasing below $T_{c}$ in CeCo$_{0.5}$Rh$_{0.5}$In$_{5}$ unlike Hg-  doped counterparts where superconductivity and magnetism coexist and the magnetic order parameter experimentally saturates in the superconducting phase.~\cite{Stock18:121}  At zero applied field, the exponent (using the temperature range discussed above) is measured to be $\beta$=0.65 $\pm$ 0.03 and $T_{N}$=2.5 K.  We note isotropic mean-field theory would predict $\beta$=0.5~\cite{Collins:book} \textcolor{black}{and the larger value derived here is likely the result of fitting an extended temperature range and not confining the analysis to temperatures as T $\rightarrow$ T$_{N}$.} No measurable change is observed for fields up to 6.5 T aligned along the crystallographic $c$ axis for either the magnetic ordering temperature, the order parameter exponent, Bragg peak position, or magnetic intensity.  We note that horizontal field measurements are limited to fields less than 6.5 T as discussed above.  With the 6.5 T field applied within the $a-b$ plane along the [1, 1, 0] axis (through rotating the horizontal field 90$^{\circ}$), $T_{N}$ is observed to increase to 2.8 K and the value of the critical exponent decreases to $\beta$=0.44 $\pm$ 0.03.  The critical exponent $\beta$ for non superconducting CeRhIn$_{5}$ is $\beta$=0.19~\cite{Stock15:114}, consistent with a two dimensional order parameter.~\cite{Bao02:65}  For slightly Hg doped CoCoIn$_{5}$, the exponent is $\beta$=0.32 as expected for three dimensional Ising order~\cite{Stock18:121}.   With the magnetic field aligned within the $a-b$ plane, as the magnetic field approaches $H_{c2}$, we conclude that the critical properties of CeCo$_{0.5}$Rh$_{0.5}$In$_{5}$ cross over to being more anisotropic in nature indicated by a decrease in the critical exponent $\beta$ (reflected by a qualitative change in the shape of the order parameter as T$\rightarrow$ T$_{N}$).  The change with exponents may also be consistent with an increase in $T_{N}$ for fields applied within the $a-b$ which can be explained by an enhancement of magnetic anisotropy that is measured to increase with applied magnetic fields~\cite{Fobes17:29} applied in the $a-b$ plane for the case of CeRhIn$_{5}$. We conclude that the critical properties indicate a crossover to increasingly anisotropic magnetism with an applied field within the $a-b$ plane.

The low temperature (in both antiferromagnetic and superconducting phases) magnetic field dependence of the $\vec{Q}$=(${1\over2}$, ${1\over2}$, ${3\over2}$) intensity for data taken from overlapping data sets from RITA and the FLEXX instruments is summarized in Fig. \ref{fig:fig1} $(c)$ for temperatures within the superconducting phase ($T$=100 mK for RITA and $T$=300 mK for FLEXX).   Note that the data has been normalized to the zero field intensity and hence the graph starts at 1.  The value of the upper and lower critical fields are $H_{c2}$=7.5 T and $H_{c1}$$\sim$10$^{-3}$ T~\cite{Majumdar03:68} respectively.  The magnetic order shows an initial increase in intensity in the vortex state below $H_{c2}$.  No strong change in intensity is found above $H_{c2}$ with a gradual decrease of intensity observable as seen with data taken on FLEXX that allowed magnetic diffraction at applied magnetic fields considerably above $H_{c2}$.  This decrease in intensity does not appear to be correlated with either of the critical fields and we now discuss this in the context of block $\uparrow\uparrow\downarrow\downarrow$ antiferromagnetism that forms at intermediate applied fields.

\subsection{$a-b$ field-induced block $\uparrow\uparrow\downarrow\downarrow$ antiferromagnetism with applied vertical fields}

Motivated by the work in Ref. \onlinecite{Raymond07:19} for CeRhIn$_{5}$ which discovered a low magnetic field induced sinusoidal order characterized by a $\vec{Q}=({1\over 2},{1\over 2},{3\over 4})$ magnetic Bragg peak, we searched for such magnetic order in CeCo$_{0.5}$Rh$_{0.5}$In$_{5}$ at low temperatures.  As displayed in Fig. \ref{fig:fig2} $(a,b)$ (taken on the WISH diffractometer and RITA2 spectrometer with a vertical $[1\overline{1}0]$ oriented field), the $\delta$L$\sim {1\over 4}$ wavevector is present and onsets at $\sim$ 2.5 T at 100 mK (Fig. \ref{fig:fig2} $c$).  \textcolor{black} {These observed peaks are resolution limited, indicative that they represent spatially long-range correlations.}  A peak at a propagation wavevector of $\delta$L=${1\over 4}$ is indicative of a further doubling of the unit cell and we interpret this as a block structure consisting of a $\uparrow\uparrow\downarrow\downarrow$ arrangement of spins along the crystallographic $c$-axis.  The combination of both $\vec{Q}=({1\over 2}, {1\over 2}, {1\over 2})$ and $({1\over 2},{1\over 2},{1\over 4})$ order is indicative of a sinusoidal modulated magnetic order present in vortex phase.  We note that for fields aligned along the crystallographic $c$-axis (measured with a horizontal field) no L=${1\over 4}$ magnetic Bragg peaks were observable at fields up to $\mu_{0}H$=6.5 T.  We speculate that this supports the observations discussed above in the context of the temperature dependence that $a-b$ plane fields enhance anisotropy to a much larger extent than when the magnetic field is applied along $c$.  \textcolor{black} {This directional dependence of anisotropy is supported by the Lande $g$-factor which is (taking the crystal field Stevens parameters from Ref. \onlinecite{Christianson04:70} for CeCoIn$_{5}$ in Table II) is 0.96 with the field oriented along $c$ and 1.95 for magnetic fields aligned within the $a-b$ plane.}  The Lande factor within the $a-b$ plane is confirmed by directional field measurements of the splitting of the spin-resonance in superconducting CeCoIn$_{5}$~\cite{Stock12:109_2} and agrees with the crystal field calculation.

\begin{figure}[t]
	\includegraphics[width=80mm,trim=2.75cm 5.75cm 3.5cm 5.75cm,clip=true]{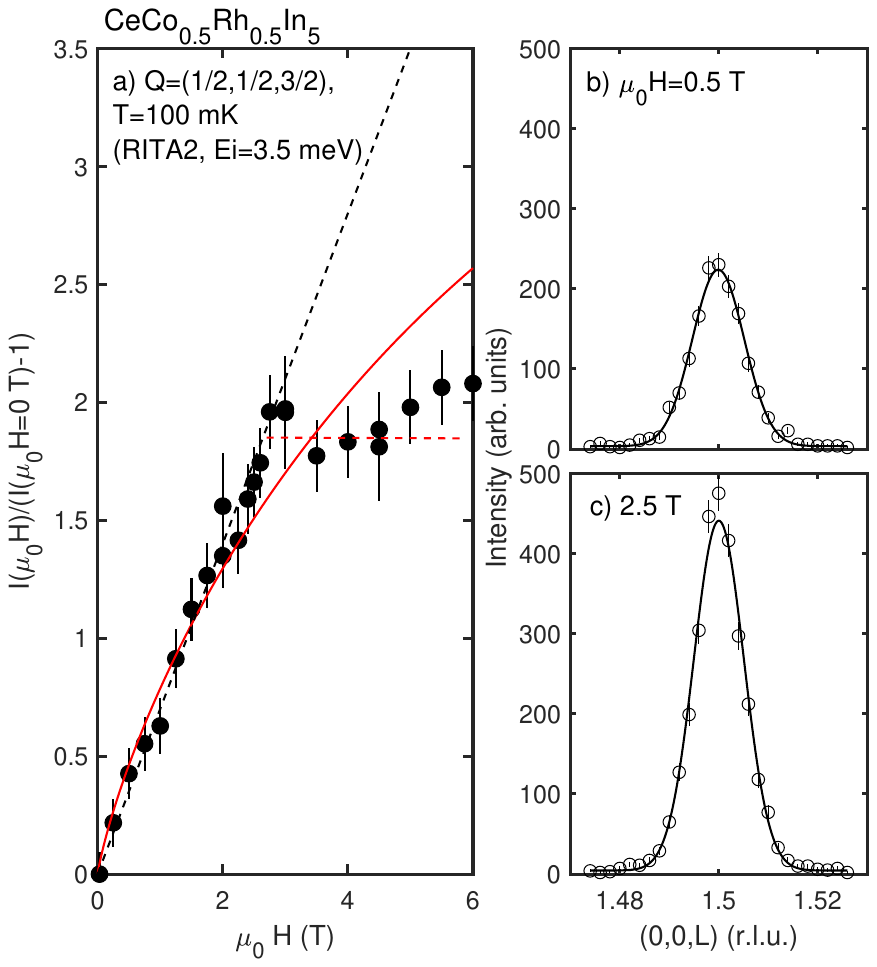}
	\caption{Magnetic field dependence using vertical fields at $\vec{Q}_{0}=({1\over 2},{1\over 2},{3\over 2})$. $(a)$ Plot of the relative change in magnetic intensity at $\vec{Q}$=(${1\over2}$, ${1\over2}$ , ${3\over2}$) in the superconducting phase as a function of magnetic field applied within the $a-b$ plane. Note that the intensity has been normalized to the $I(\mu_{0}H=0$ T$)-1$ in comparison to Figure 2 where it was normalized by by just $I(\mu_{0}H=0$ T).  The dashed black line is a linear fit and the red dashed curve is a fit to $I(\mu_{0}H) \propto H/H_{c2} \ln (\theta H_{c2}/H)$, for proximate spin density wave and superconducting orders.  The average intensity above $\sim$ 2 T is denoted by the horizontal dashed red line.  $(b-c)$ illustrate two scans along L, taken on RITA, showing the increase in intensity and the commensurate $\vec{Q}$=(${1\over2}$, ${1\over2}$, ${3\over2}$) order.} \label{fig:fig3}
\end{figure}

\subsection{Interplay of $a-b$ field induced vortex state and $\uparrow\downarrow$ antiferromagnetism with applied vertical fields}

We now discuss the scaling of the magnetic intensity of the dominant $\vec{Q}$=(${1\over2}$, ${1\over2}$, ${3\over2}$) magnetic Bragg peak with the magnetic field applied within the $a-b$ plane using $[1\overline{1}0]$ oriented vertical magnetic fields.  Fig. \ref{fig:fig3} $(a)$ plots the intensity as a function of applied field at $T$=100 mK taken on RITA2 with the magnetic intensity rising linearly with field (dashed line) for small applied magnetic fields and then a more gradual increase or even saturation at fields larger than $\sim$ 2 T.  Note that the intensity has been scaled by the zero field intensity and normalized such that the graphs starts from 0 to facilitate fitting of the field dependence and comparison to models previously applied to the cuprates.  The integrated intensity was extracted from Gaussian fits with representative curves shown in Figs. \ref{fig:fig3} $(b-c)$ which display this increase through elastic scans along $L$.  Scans both along the (H,H,0) and (0,0,L) directions find no observable change in peak position over this field range below $H_{c2}$.  The linear increase is not expected for this geometry, based on magnetic and crystalline structure alone, given that the two-fold symmetry is preserved~\cite{Stock19:100} when the field is applied along this direction. This would predict any change with magnetic field should be even, hence quadratic in field $\sim|H|^{2}$ to leading order.  It is also not expected for purely magnetic effects such as antiferromagnetism near dopant or impurity sites.~\cite{Kohno99:68,Stock05:74,Stock09:79}  The linear $\sim |H|$ change implies a coupling to a quantity that changes linearly with applied magnetic field.  One such candidate is the superconducting order parameter $|\psi|^{2}$ which is reduced on the application of a magnetic field owing to the introduction of vortex cores.  In the presence of vortices, the superconducting volume fraction is suppressed and this could cause an increase in the magnetic order parameter $|\phi|^{2}$ coupled to the superconducting order parameter $|\psi|^{2}$.  Following the arguments of Ref. \onlinecite{Khaykovich02:66} for an analogous result in La$_{2}$CuO$_{4+y}$, since the magnetic correlation length is large (defined by resolution limited magnetic Bragg peaks), the magnetic order is averaged over regions outside the vortices and scales with the magnetic field.  

Such a situation was considered for proximate spin density wave and superconducting orders in the cuprate superconductors in Ref. \onlinecite{Demler01:87,Zhang02:66}, motivated by experimental neutron inelastic scattering work.~\cite{Lake01:291,Lake05:4}   Given that the number of vortices above $H_{c1}$ scales linearly with field, it was initially suggested that enhanced magnetic signal originated from spins in the vortex cores~\cite{Hu02:63,Vaknin00:329,Arovas97:79}. However, such a description was found to be inconsistent with the expected charge redistribution.  It is also inconsistent with the resolution limited nature of the magnetic peak indicative of a correlation length of at least $\sim$ 500 \AA\ that exceeds the dimensions of a vortex core which is expected to be on the order of several unit cells or $\sim$ 10 \AA.   

By considering the response of the superconducting order parameter outside the vortex (expected to scale with distance as $\sim 1/2x^{2}$), the theoretical prediction for the scaling of magnetic intensity with field after performing a spatial two dimensional integral is $\sim H/H_{c2} \ln (\theta H_{c2}/H)$ to leading order.~\cite{Demler01:87}  A fit to this relation is shown by the solid curve in Fig. \ref{fig:fig3} $(a)$ with $H_{c2}$ fixed at 7.5 T  and $\theta$=6.8 $\pm$ 1.0.  The value of $\theta$ is sensitive to the vortex core geometry and is predicted to be $\sim$ 3 for square or triangular vortices.  The linear scaling of the magnetic Bragg peak intensity (proportional to the magnetic order parameter squared $|\phi|^{2}$) at low magnetic fields below $H_{c2}$ is consistent with this prediction for the change in the magnetic response near a quantum critical point to superconducting and spin density wave order.  Presumably here the quantum critical point would be $x$ $\sim$ 0.4 displayed in Fig. \ref{fig:phase_diagram}. However, and we emphasize, that while the logarithmic form captures the initial linear rise in magnetic intensity for small applied fields observed in our experiment, the strong deviation from the fit at $\sim$ 2 T combined with the anomalously large fitted $\theta$ indicates that the logarithmic correction may not be appropriate for this situation over the field range applied.  This could be the result of experimental conditions being, in fact, located far in parameter space from any quantum critical point or the onset of a new order parameter.

\begin{figure}[t]
	\includegraphics[width=93mm,trim=0cm 5cm 0cm 6cm,clip=true]{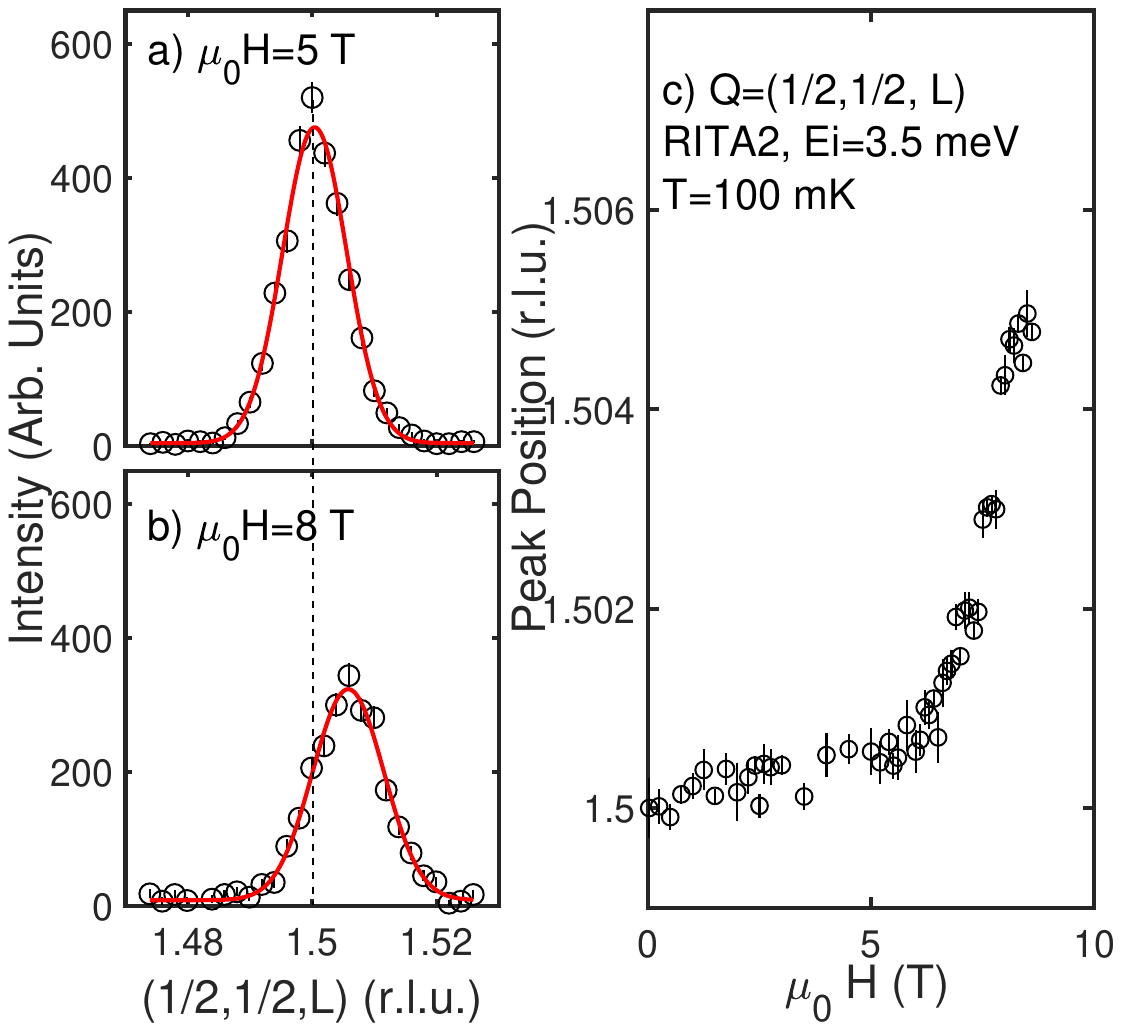}
	\caption{Magnetic diffraction near $\vec{Q}_{0}=({1\over 2},{1\over 2},{1\over 2})$ in vertical fields. $(a-b)$ Scans along $({1\over2}, {1\over2}, L)$ in the vortex and normal phases.  $(c)$ plot of the peak position along $L$ as a function of magnetic field using the RITA2 cold triple-axis spectrometers at T=100 mK. } \label{fig:fig4}
\end{figure}

It is interesting that the field where a large deviation is observed from this low-field theoretical description is near the same field range where we observe the onset of $\uparrow\uparrow\downarrow\downarrow$ order evidenced by a magnetic Bragg peak at $L={1\over 4}$ positions.  This is suggestive that block $\uparrow\uparrow\downarrow\downarrow$ order interrupts the competition between magnetic $|\phi|^{2}$ and superconducting $|\psi|^{2}$ orders.  As shown in Fig. \ref{fig:fig1} $(c)$, the magnetic Bragg peak intensity decreases gradually to fields well above $H_{c2}$.  For small applied fields within the $a-b$ plane, we conclude that there is a coupling between magnetic and superconducting order parameters.  For large field induced anisotropies, the response differs owing to the presence of anisotropic block magnetism.  This coupling is anisotropic and not experimentally observable for fields applied along the crystallographic $c$ axis.

\subsection{$a-b$ plane induced normal state incommensurate magnetism with vertical fields}

\begin{figure}[t]
	\includegraphics[width=93mm,trim=0cm 4.5cm 0cm 5cm,clip=true]{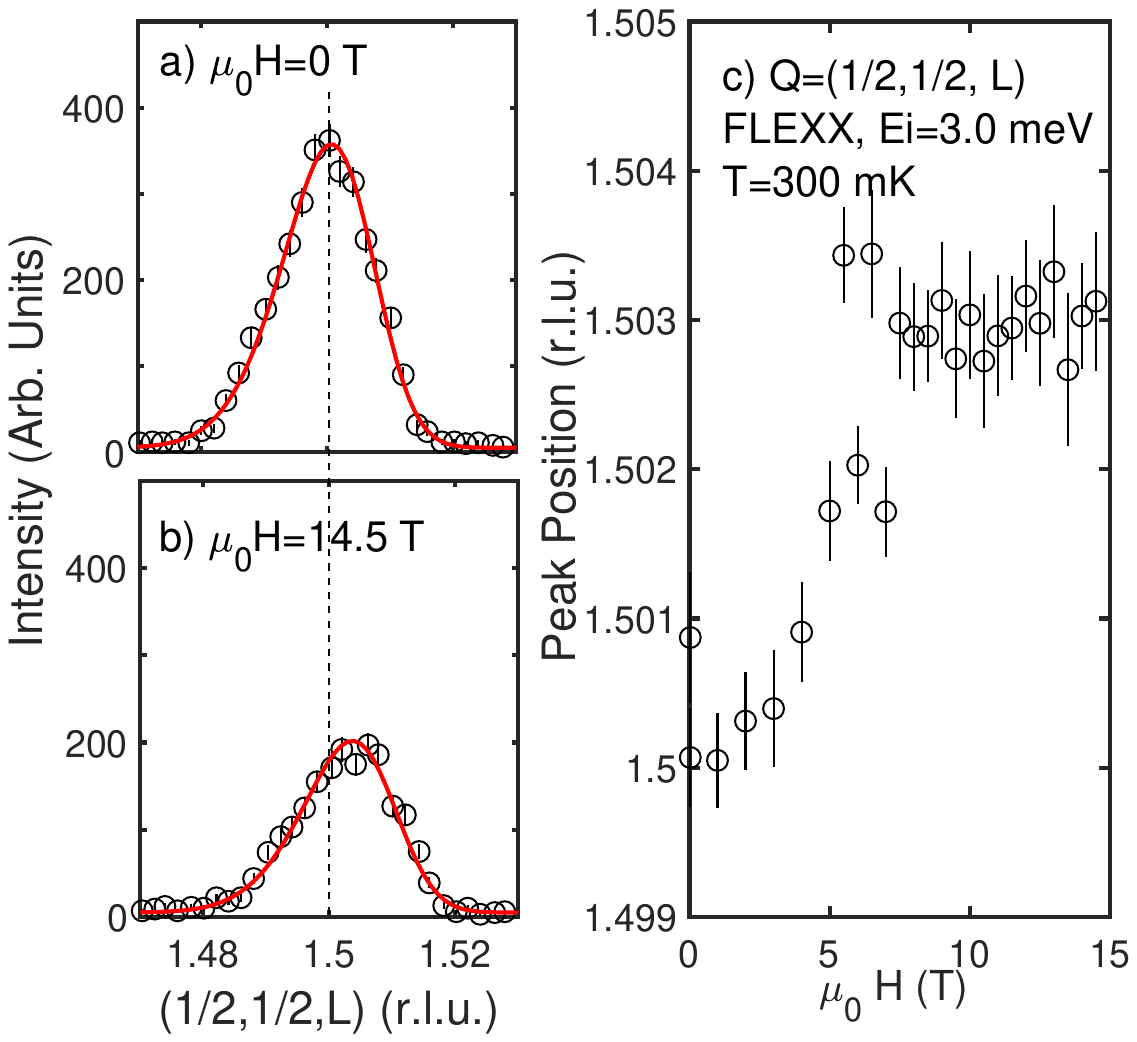}
	\caption{Magnetic diffraction near $\vec{Q}_{0}=({1\over 2},{1\over 2},{1\over 2})$ in vertical fields. $(a-b)$ Scans along $({1\over2}, {1\over2}, L)$ in the vortex and normal phases.  $(c)$ plot of the peak position along $L$ as a function of magnetic field using the FLEXX cold triple-axis spectrometers at T=300 mK.} \label{fig:figHZB}
\end{figure}

In Figs. \ref{fig:fig4} and \ref{fig:figHZB}, we investigate the magnetic propagation wavevector as a function of magnetic field and in particular near the transition from the vortex state to the normal state with increasing field.  The comparatively low value of $H_{c2}$ that exists for CeCo$_{0.5}$Rh$_{0.5}$In$_{5}$ allows the superconducting order parameter to be completely suppressed with magnetic fields that are available at current neutron sources.  We present two datasets from two different instruments owing to limitations of available magnetic fields on each beamline.  Fig. \ref{fig:fig4} was measured at RITA2 at $T$=100 mK with magnetic fields constrained below 9 T.  Given $H_{c2}$=7.5 T, we pursued higher magnetic fields at T=300 mK at the FLEXX cold triple-axis spectrometer where magnetic fields of 14.5 T could be reached which were well in the normal state allowing us to establish trends at large magnetic fields.  

Figure \ref{fig:fig4} $(a-b)$ illustrates scans through the magnetic Bragg peak in the $(a)$ vortex and $(b)$ at fields just within the normal phase.  A shift in the position in $L$ is observed away from the commensurate $L={1\over2}$ position while scans along the (H,H,0) direction found no observable change of the position in H with field.  Fig. \ref{fig:fig4} $(c)$ shows that this change in the $L$ position occurs near $H_{c2}$ based on fits to scans along $L$ to a Gaussian.  To study the dependence into the normal state we used the FLEXX cold triple-axis instrument were applied fields of $\mu_{0}H$=14.5 T could be reached.  These are illustrated in Fig. \ref{fig:figHZB}.  Fig. \ref{fig:figHZB} $(a)$ and $(b)$ illustrate scans along $L$ in the superconducting and normal phases.  Because of the asymmetric momentum resolution, the following function was used to extract the peak position along $L$ as a function of applied magnetic field,

\begin{equation}\label{eq:skewed}
	I(L) \propto \left\{1+\erf{\left[\frac{\gamma(L-L_0)}{\sigma\sqrt{2}}\right]}\right\}\exp{-(L-L_0)^2 \over 2\sigma^2}. \nonumber
\end{equation}

\noindent  Here, $\gamma$ is a parameter to fit the peak assymmetry and $\sigma$ and $L_{0}$ are the standard deviations and position of the peak.  The peak position is plotted in Fig. \ref{fig:figHZB} $(c)$ which illustrates a slightly smaller shift to incommensurate $L$ and at smaller fields.  This could be the result of greater field inhomogeneities and also the higher temperature of the measurement in comparison to the RITA2 data.  Nevertheless, Fig. \ref{fig:figHZB} $(c)$ illustrates the shift from commensurate to incommensurate magnetism and also the saturation in the normal state.  

The incommensurate shift observed is very small, particularly in the context of the propagation wave vector for CeRhIn$_{5}$ which is 0.297 r.l.u along $L$ and we discuss experimental considerations here.  Kinematically, the observed shift we observe on FLEXX and RITA2 corresponds to an angular shift primarily along $2\theta$ (or A4, using triple-axis angle labeling convention~\cite{Shirane:book}) of $\sim 0.2^{\circ}$ and less than 0.1$^{\circ}$ in sample rotation (or A3).  If the sample was twisting in the beam due to, for example, forces on the sample in the Meissner effect, we would expect a change in the sample rotation which is not observed here.  This can also be discounted as measurements of nuclear Bragg peaks did not show a shift in angle with applied field.  The field in this experiment was only altered when the sample was heated to the normal state as discussed above in the experimental section.  

\textcolor{black} {We now discuss the incommensurate propagation wavevector measured at large magnetic fields.  For a transition from commensurate magnetic order to incommensurate characterized by a wave vector of $q$,  we would expect domains resulting in incommensurate peaks at $\pm \delta q$.  For the position scanned near $({1\over2},{1\over2},{3\over2})$ above using RITA and FLEXX, this would result in magnetic peaks originating from  (0,0,1) $+ \delta q$ and (1,1,2) $- \delta q$.  This would \textit{not} result in a single peak that is shifted, but rather a splitting in the Bragg peak.  This is not observed experimentally in Figs. \ref{fig:fig4}. and \ref{fig:figHZB}.  It is possible that the magnetic field suppresses domains resulting in a single domain with a fixed propagation wavevector $q$.  We note that such single domain states are present in some materials such as iron-based langasite~\cite{Stock19:100,Stock11:83,Qureshi20:102,Marty08:101}.  Furthermore, CeRhIn$_{5}$ displays such preferential magnetic domains on cooling through T$_{N}$ which allows the helical magnetic structure to be uniquely identified with spherical neutron polarimetry.~\cite{Stock15:114}}   

We therefore suggest that the magnetism in CeCo$_{0.5}$Rh$_{0.5}$In$_{5}$ transitions from commensurate magnetism coexisting with superconductivity to incommensurate magnetism at high fields and low temperatures in the normal phase.

\subsection{Momentum and energy broadened excitations at zero field}

\begin{figure}[t]
	\includegraphics[width=100mm,trim=4cm 5.5cm 1cm 5cm,clip=true]{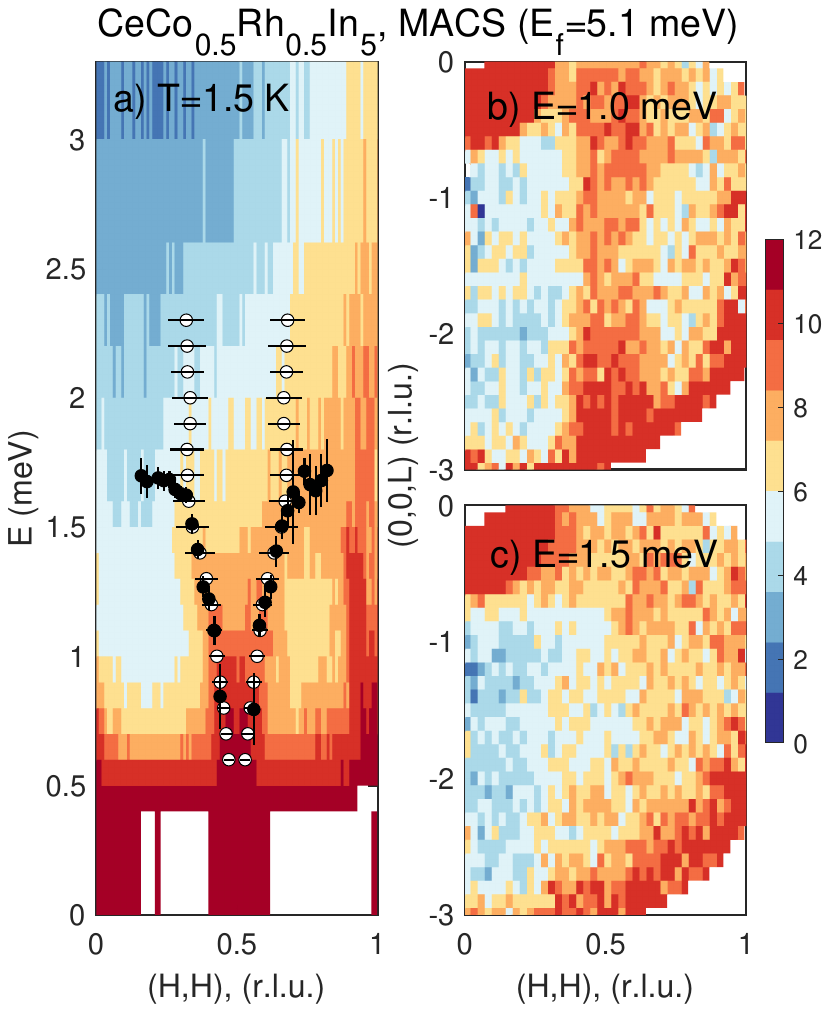}
	\caption{Magnetic spectroscopy at zero applied field probing the low-energy magnetic excitations measured on MACS (NIST).  $(a)$ displays a constant momentum slice integrated over L=[-3,-1].  The solid (open) circles are peak positions fitted to the magnetic excitations in CeRhIn$_{5}$ (Ref. \onlinecite{Stock15:114}) taken from constant momentum (energy) scans.  The data illustrate comparatively momentum and energy broadened excitations dispersing from the commensurate $({1\over 2},{1 \over 2})$ position. $(b,c)$ show constant energy slices at E=1.0 and 1.5 meV.  The rod of scattering along $L$ is indicative of two dimensional magnetic fluctuations.} \label{fig:excitations}
\end{figure}

Finally, we investigate the magnetic excitations below $T_{N}$ in CeCo$_{0.5}$Rh$_{0.5}$In$_{5}$ at zero applied field and compare the results to parent antiferromagnetic CeRhIn$_{5}$.  The magnetic dynamics at this temperature and field correspond to excitations from a commensurate (${1\over 2}$,${1\over 2}$,${1\over 2}$) order which differs from the incommensurate order found in the high-field normal phase of CeCo$_{0.5}$Rh$_{0.5}$In$_{5}$ and also at ambient conditions in CeRhIn$_{5}$.  While CeRhIn$_{5}$ has been reported to display well defined, and underdamped, magnetic excitations~\cite{Das14:113}, these excitations have been shown to coexist with an energy and momentum broadened continuum~\cite{Stock15:114} which is indicative of excitations polarized within the $a-b$ plane being unstable.  The magnetic excitations in the commensurate phase of  CeCo$_{0.5}$Rh$_{0.5}$In$_{5}$ are displayed in Fig. \ref{fig:excitations}.  A constant momentum slice is displayed in Fig. \ref{fig:excitations} $(a)$ integrated over the range of L=[-3,-1] and primarily sensitive to excitations polarized within the $a-b$ plane.  The solid and empty-white points are from CeRhIn$_{5}$ with the solid points the peak position of underdamped magnetic excitations obtained from constant-momentum cuts and the open circles from constant energy cuts.  The open circles differ from the filled at higher energies owing to the presence of a momentum and energy broadened continuum of excitations present at higher energies in CeRhIn$_{5}$.  

The magnetic excitations emanate from the (${1\over 2}$,${1\over 2}$) position and unlike the case in incommensurate CeRhIn$_{5}$, are not well defined and extended in momentum and energy.  \textcolor{black} {While the magnetic scattering follows the dispersion of CeRhIn$_{5}$, the scattering is less well defined in energy and momentum and this can be seen at small values of (H,H) in Fig. \ref{fig:excitations} $(a)$.} Figs. \ref{fig:excitations} $(b,c)$ illustrate constant E=1.0 and 1.5 meV slices showing that the while the scattering are correlated around (H,H)= (${1\over 2}$,${1\over 2}$), it forms a rod of scattering along $L$. This is indicative of two dimensional correlations analogous to that observed in CeRhIn$_{5}$ over the same energy range. \textcolor{black}{While the magnetic scattering forms rods consistent with two-dimensional correlations, the constant energy slices in Figs. \ref{fig:excitations} $(b,c)$ to not display two clear peaks emanating from the (H,H)= (${1\over 2}$,${1\over 2}$) as observed in antiferromagnetic CeRhIn$_{5}$ and indicative of dispersive underdamped spin-waves. The lack of well defined and underdamped spin excitations  indicates that with increased substitution towards superconducting CeCoIn$_{5}$, in-plane excitations become unstable shortening lifetimes and dynamic correlation lengths.}  We note that this is consistent with the lack of observable in-plane excitations in superconducting CeCoIn$_{5}$ and Hg substituted variants.~\cite{Stock08:100,Stock18:121}  

The lack of any clear magnetic fluctuations that resemble antiferromagnetic parent CeRhIn$_{5}$ is in contrast to the observations in chemically substituted insulators like K(Mn,Co/Zn)F$_{3}$~\cite{Buyers72:5,Buyers71:27} or Rb$_{2}$(Mn,Co)F$_{4}$~\cite{Ikeda80:49}.  In these compounds experiments observed the spin excitations having modified structures of the underlying Mn- and Co- parent constituents~\cite{Buyers73:6}. We therefore speculate the momentum and energy broadening of the magnetic excitations in CeCo$_{0.5}$Rh$_{0.5}$In$_{5}$ originates from a coupling to itinerant electronic carriers.  Such a coupling is consistent with the competing magnetic and superconducting order parameters discussed above in the context of the diffraction results.

\section{Discussion and Conclusions}

We have observed incommensurate magnetic order at high magnetic fields in the normal phase which competes with commensurate magnetism in the superconducting and vortex phases in CeCo$_{0.5}$Rh$_{0.5}$In$_{5}$.  The incommensurate order observed here differs from that previously reported in field induced density wave phases in CeCoIn$_{5}$~\cite{Kenzelmann10:104,Blackburn10:105} which is also stabilized through Nd doping Ce(Co,Rh)In$_{5}$~\cite{Mazzone17:3,Rosa17:114}.   The ``Q-phase", reported in these materials, exists only in the superconducting phase and is furthermore characterized by a propagation wavevector that is incommensurate within the crystallographic $a-b$ plane.~\cite{Murphy02:65,Young07:98,Kenzelmann08:321,Blackburn10:105}  In CeCo$_{0.5}$Rh$_{0.5}$In$_{5}$, the magnetic order is commensurate within the superconducting phase and becomes incommensurate along $L$ in the normal phase.  The incommensurate change is small in comparison to the incommensurate L=0.297 order found in CeRhIn$_{5}$ indicating a long wavelength modulated magnetic phase competes with unconventional superconductivity.  The long wavelength incommensurate order that we observe is a property of the normal state as we observe little change in this ordering wavevector for fields up to 14.5 T in comparison to the critical field of 7.5 T.  Incommensurate order competing with superconductivity is a common trend and has also been found in, for example, iron based systems through doping studies~\cite{Rodriguez11:83,Rodriguez84:13,Stock17:95} as well as the heavy fermion YbRh$_{2}$Si$_{2}$~\cite{Schuberth16:351,Stock12:109} (though neutron scattering has only been done in the normal non superconducting phase).  

It is important to note that changes found experimentally here are not observable when the field is applied along the crystallographic $c$-axis.  As noted above and also directly measured by neutron spectroscopy in Ref. \onlinecite{Fobes17:29}, when the field is applied within the $a-b$ plane there is an increase in the energetic gap of the magnetic excitations indicative of an increase in magnetic anisotropy.  This is consistent with the increase in $T_{N}$ observed for this field geometry in Fig. \ref{fig:fig1} $(b)$ as enhanced anisotropy would favor magnetic order at higher temperatures.  Assuming a connection between an increase in $T_{N}$ with applied field, it would imply that applying the magnetic field along the $c$-axis has comparatively little effect on anisotropy.  However, unlike the case of $a-b$ field geometries, no neutron spectrosocpy data currently exists for fields oriented along the crystallographic $c$-axis which would require a horizontal magnet.  Given the response observed in the ordering wavevector and the magnetic intensity here, to conclusively connect the field induced anisotropy with this effect would require further spectroscopy work on new instrumentation with horizontal magnetic fields.  Nevertheless, we speculate that the anisotropic response observed here underscores the importance of magnetic anisotropy in analyzing magnetic and superconducting order parameters in the `115' Ce-based compounds.


\textcolor{black} {It is important to compare the results presented here to CeRhIn$_{5}$ under applied pressure.  As measured in Ref. \onlinecite{Llobet04:69}, with increasing pressure incommensurate (helical) magnetism coexists with superconductivity for large pressures and abrupt changes in the Fermi surface have been reported under pressure (Ref. \onlinecite{Shishido06:378}), analogous to the discussion and situation illustrated in Fig. \ref{fig:phase_diagram}.  In comparison to the field dependent magnetism discussed here for CeCo$_{0.5}$Rh$_{0.5}$In$_{5}$, a slight change in the incommensurate wavevector increasing towards the commensurate L$\sim {1\over3}$ position is observed in the pressure induced superconducting state in Ref. \onlinecite{Llobet04:69} Figure 2 $(b)$.  This maybe analogous to the situation here where superconducting seems to favor commensurate magnetism, becoming incommensurate in the normal state.}

Assuming a coupling between magnetic and electronic parameters evidenced by competing static orders and momentum-energy broadened excitations, the data presented here could illustrate a competition on the Fermi surface between incommensurate magnetic order and superconductivity.~\cite{Kawamura07:76}  Associating  the magnetic propagation vector with a nesting on the Fermi surface would imply that part of the Fermi surface is involved and destroyed in the superconducting phase.  Such scenarios have been proposed for CeRhIn$_{5}$~\cite{Alvarez07:98,Chen06:97} and previously suggested to explain the interplay between magnetism and superconductivity in URu$_{2}$Si$_{2}$~\cite{Maple86:56} as well as the field dependence in the bilayer cuprates~\cite{Haug09:103}.  This has also been suggested theoretically~\cite{Alvarez07:98} by applying mean-field theory.   Given that the incommensurate wave vector in the normal state is along the (0,0,L) direction, two dimensional antiferromagnetic in-plane correlations favor superconductivity as suggested in Ref. \onlinecite{Chang19:99}.  We note that the close vicinity of incommensurate magnetism may be connected with the non-Fermi liquid behavior seen at large fields in CeCoIn$_{5}$.~\cite{Paglione03:91,Paglione06:97}

It is interesting to note the instability of magnetic excitations originating from the commensurate magnetically ordered phase at zero applied field in CeCo$_{0.5}$Rh$_{0.5}$In$_{5}$.  These dynamics were sampled at large values of momentum transfer along the crystallographic $c$-axis and therefore are sensitive to magnetic excitations polarized within the $a-b$ plane.  We note that no observable magnetic excitations polarized within the $a-b$ plane were observable in superconducting CeCoIn$_{5}$ or Hg substituted counterparts.  Therefore, in-plane magnetic excitations associated with commensurate magnetic order appear to be unstable on substitution of Co and on entering the superconducting phase.  This further illustrates the importance of anisotropy in the magnetic response. 

In summary, we report the critical properties of magnetic order near Co substitution values which separate incommensurate magnetism from superconductivity in CeCo$_{0.5}$Rh$_{0.5}$In$_{5}$.  We find a scaling of the antiferromagnetic order parameter consistent with the suppression of the superconducting order parameter away from the vortex cores at small fields and have noted the importance of block ($L={1\over4}$) order.  On entering the normal state, the commensurate response is replaced by a weakly incommensurate along $c$ propagation wave vector.  These effects are not observable for magnetic fields oriented along the $c$-axis which we suggest maybe related to the field dependence of the magnetic anisotropy.  The results suggest a common point in the `115' phase diagram where there is an interplay between superconductivity and incommensurate magnetism.  

\section{Acknowledgements}

This work was funded by the STFC and the EPSRC.  Work at the University of Maryland is supported by the Gordon and Betty Moore Foundations EPiQS Initiative through Grant No. GBMF4419, and the National Institute of Standards and Technology Cooperative Agreement 70NANB17H301


%

\end{document}